\documentclass[twocolumn,preprintnumbers]{revtex4}

\usepackage{graphicx}
\usepackage{epsf}
\usepackage{amsmath,amssymb}
\usepackage{color}

\usepackage{ulem}

\begin{document}
\preprint{KUNS-2758, NITEP 17}
\title{First microscopic coupled-channel calculation of $\alpha$
inelastic cross sections on $^{16}$O}

\author{Yoshiko Kanada-En'yo}
\affiliation{Department of Physics, Kyoto University, Kyoto 606-8502, Japan}
\author{Kazuyuki Ogata} 
\affiliation{Research Center for Nuclear Physics (RCNP), Osaka University,
  Ibaraki 567-0047, Japan}
\affiliation{Department of Physics, Osaka City University, Osaka 558-8585,
  Japan}
\affiliation{
Nambu Yoichiro Institute of Theoretical and Experimental Physics (NITEP),
   Osaka City University, Osaka 558-8585, Japan}
\begin{abstract}
The $\alpha$ inelastic scattering on $^{16}$O is investigated with the
coupled-channel calculation using the $\alpha$-nucleus coupled-channel potentials, which are
microscopically derived by folding the
the Melbourne $g$-matrix $NN$ interaction with the $^{16}$O and $\alpha$
densities.
The matter and transition densities of $^{16}$O are calculated by a microscopic structure model of
the variation after the spin-parity projections
combined with the generator coordinate method of $^{12}$C+$\alpha$ in the framework of the
antisymmetrized molecular dynamics.
The calculation reproduces the observed elastic and inelastic cross sections
at incident energies of $E_\alpha=104$ MeV, 130 MeV, 146 MeV, and 386 MeV.
The coupled-channel effect on the cross sections is also discussed.
\end{abstract}
\maketitle

\section{Introduction}

The $\alpha$ scattering has been used for study of isoscalar (IS) monopole and dipole
excitations in nuclei.
The inelastic cross sections have been analyzed by
reaction model calculations to determine the strength functions in a wide range of excitation energy covering the giant resonances
\cite{Harakeh-textbook}.
The $\alpha$ scattering is also a good tool to probe cluster states
because these have  generally strong IS
monopole and dipole transition strengths and can be populated  by
the $\alpha$ scattering reaction
\cite{Kawabata:2005ta,Yamada:2011ri,Chiba:2015khu}.
Indeed, the $(\alpha,\alpha')$ reaction experiments have been intensively performed
to investigate cluster structures of excited states in light nuclei such as $^{12}$C and $^{16}$O
recently.

For the study of cluster structures in $^{12}$C,
the $^{12}$C($\alpha,\alpha'$) reaction has been investigated
with reaction models \cite{John:2003ke,Ohkubo:2004fu,Takashina:2006pc,Khoa:2007as,Takashina:2008zz,Itoh:2011zz,Ito:2018opr,Adachi:2018pql},
but many of the reaction calculations encountered the overshooting problem  of the $0^+_2$ cross sections.
Also for  $sd$-shell nuclei
such as $^{16}$O, $\alpha$ scattering experiments have reported
the similar overshooting problem of the $0^+$ cross sections
in the reaction model analysis \cite{Adachi:2018pql}.
Recently, Minomo and one of the authors (K.~O.)
have carried out microscopic coupled-channel calculation and succeeded
in reproducing the $0^+_2$ cross sections of the $^{12}$C($\alpha,\alpha'$)
reaction with no adjustable parameter \cite{Minomo:2016hgc}.
In the study, $\alpha$-nucleus CC potentials are constructed by folding
the Melbourne $g$-matrix effective $NN$ interaction \cite{Amos:2000}
by a phenomenological matter density of $\alpha$ and the matter and transition densities of $^{12}$C obtained with the resonating group method \cite{Kamimura:1981oxj}.
In our previous paper  \cite{Kanada-En'yo:a12C}, we have applied the $g$-matrix folding model to the same reaction
and reproduced the cross sections of the
$0^+_{2,3}$, $1^-_1$, $2^+_{1,2}$, and $3^-_1$ states of $^{12}$C
with the transition density obtained by the antisymmetrized molecular dynamics (AMD)
 \cite{KanadaEnyo:1995tb,KanadaEnyo:1995ir,KanadaEn'yo:1998rf,KanadaEn'yo:2001qw,KanadaEn'yo:2012bj},
which is a microscopic structure model beyond the cluster models.
These works indicate that, if reliable transition densities are available
from structure model calculations,
the approach of the $g$-matrix folding model
can be a useful tool to investigate cluster states by the ($\alpha,\alpha'$) reaction.

In the structure studies of $^{16}$O,
a variety of cluster structures
such as the  $4\alpha$-tetrahedral,
$^{12}$C+$\alpha$, and a $4\alpha$-cluster gas state have been suggested by the cluster models
\cite{Yamada:2011ri,wheeler37,dennison54,brink70,Suzuki:1976zz,Suzuki:1976zz2,fujiwara80,Libert-Heinemann:1980ktg,bauhoff84,Descouvemont:1987uvu,Descouvemont:1991zz,Descouvemont:1993zza,Fukatsu92,Funaki:2008gb,Funaki:2010px,Horiuchi:2014yua,Bijker:2014tka,Bijker:2016bpb}.
Recently, the experimental studies of $^{16}$O have been performed
by the $^{16}$O($\alpha,\alpha'$) reaction \cite{Wakasa:2006nt,Adachi:2018pql}. In Ref.~\cite{Wakasa:2006nt}, the
 $0^+_4$ state at 13.6 MeV has been discussed in relation with the $4\alpha$-gas state
with the reaction model analysis using phenomenological CC potentials. In the study, the $\alpha$-scattering cross sections are naively assumed to scale the IS monopole strengths.

However, no microscopic CC calculation of the $^{16}$O$(\alpha,\alpha')$
reaction was performed so far, mainly because of the
theoretical difficulty of microscopic structure models in description of  $^{16}$O.
For instance, a well-known problem is that microscopic cluster models largely overshoot
the excitation energy of the $K^\pi=0^+_2$ band.
Recently, one of the authors (Y.~K-E.) investigated the cluster structures of $^{16}$O
\cite{Kanada-En'yo:2013dma,Kanada-Enyo:2017ers,Kanada-Enyo:2019hrm}
with the AMD.
She has performed the variation after spin-parity projections (VAP)
combined with the generator coordinate method (GCM) of the $^{12}$C+$\alpha$ cluster
in the AMD framework,  which we called the VAP+GCM.
The VAP+GCM calculation qualitatively described
the energy spectra of $^{16}$O and obtained the
$0^+_2$, $2^+_1$, $4^+_1$, $1^-_2$, and $3^-_2$ states
in the $^{12}$C+$\alpha$ bands, and also the
$3^-_1$ and $4^+_2$ states in the $4\alpha$-tetrahedral ground band.
Moreover,  it predicts the $4\alpha$-gas state as the $0^+_5$ state near the $4\alpha$ threshold energy.

In this paper, we apply the $g$-matrix folding model to
the $^{16}$O$(\alpha,\alpha')$ reaction using the matter and transition densities
calculated with the VAP+GCM in a similar way to in our previous work on  the $^{12}$C$(\alpha,\alpha')$ reaction
\cite{Kanada-En'yo:a12C}.
The present work is the first microscopic CC calculation of the
 $^{16}$O$(\alpha,\alpha')$ reaction that is based on  the
microscopic $\alpha$-nucleus CC potentials derived with
the $g$-matrix folding model.
The calculated cross sections are compared with the observed data at incident energies
of $E_\alpha=104$ MeV, 130 MeV, 146 MeV, and 386 (400) MeV
\cite{Adachi:2018pql,Wakasa:2006nt,Hauser:1969hkb,Knopfle1975,Harakeh:1976rtd}.
The IS monopole and dipole transitions to the $0^+_{2,3,4,5}$ and $1^-_1$ states are focused.
We try to answer the following questions.
Can the microscopic reaction calculation
describe the $\alpha$ scattering cross sections?
Does the overshooting problem of the monopole strength exist?
Is the scaling law of the $\alpha$ scattering cross sections and the IS monopole transition strength satisfied?

The paper is organized as follows.
Section~\ref{sec:structure-model} describes the structure calculation of $^{16}$O
with the VAP+GCM, and Sec.~\ref{sec:scattering} discusses the $^{16}$O$(\alpha,\alpha')$ scattering investigated
with the microscopic CC calculation.
Finally, a summary is given in Sec.~\ref{sec:summary}.

\section{Structure calculation of $^{16}$O with VAP+GCM}\label{sec:structure-model}
\subsection{Wave functions of $^{16}$O}
The wave functions of $^{16}$O are those obtained  by the
variation after spin-parity projections (VAP) \cite{KanadaEn'yo:1998rf} combined with the $^{12}$C+$\alpha$
GCM in the AMD framework, which we called the VAP+GCM \cite{Kanada-Enyo:2017ers}.
As shown in Ref.~\cite{Kanada-Enyo:2019hrm},
the VAP+GCM calculation reasonably reproduces the energy spectra and transition strengths of $^{16}$O,
and obtains various cluster structures such as the $4\alpha$ and $^{12}$C+$\alpha$ cluster structures.
For the details of the formulation of the structure calculation and the resulting structures and band assignments in $^{16}$O,
the reader is referred to Refs.~\cite{Kanada-En'yo:2013dma,Kanada-Enyo:2017ers,Kanada-Enyo:2019hrm}.
Using the VAP+GCM wave functions, the transition strengths, matter and transition densities, and form factors are
calculated. The definitions of these quantities are given in Refs.~\cite{Kanada-En'yo:a12C,Kanada-Enyo:2019hrm}.

\subsection{Excitation energies and radii}

\begin{figure}[!h]
\begin{center}
\includegraphics[width=8.6cm]{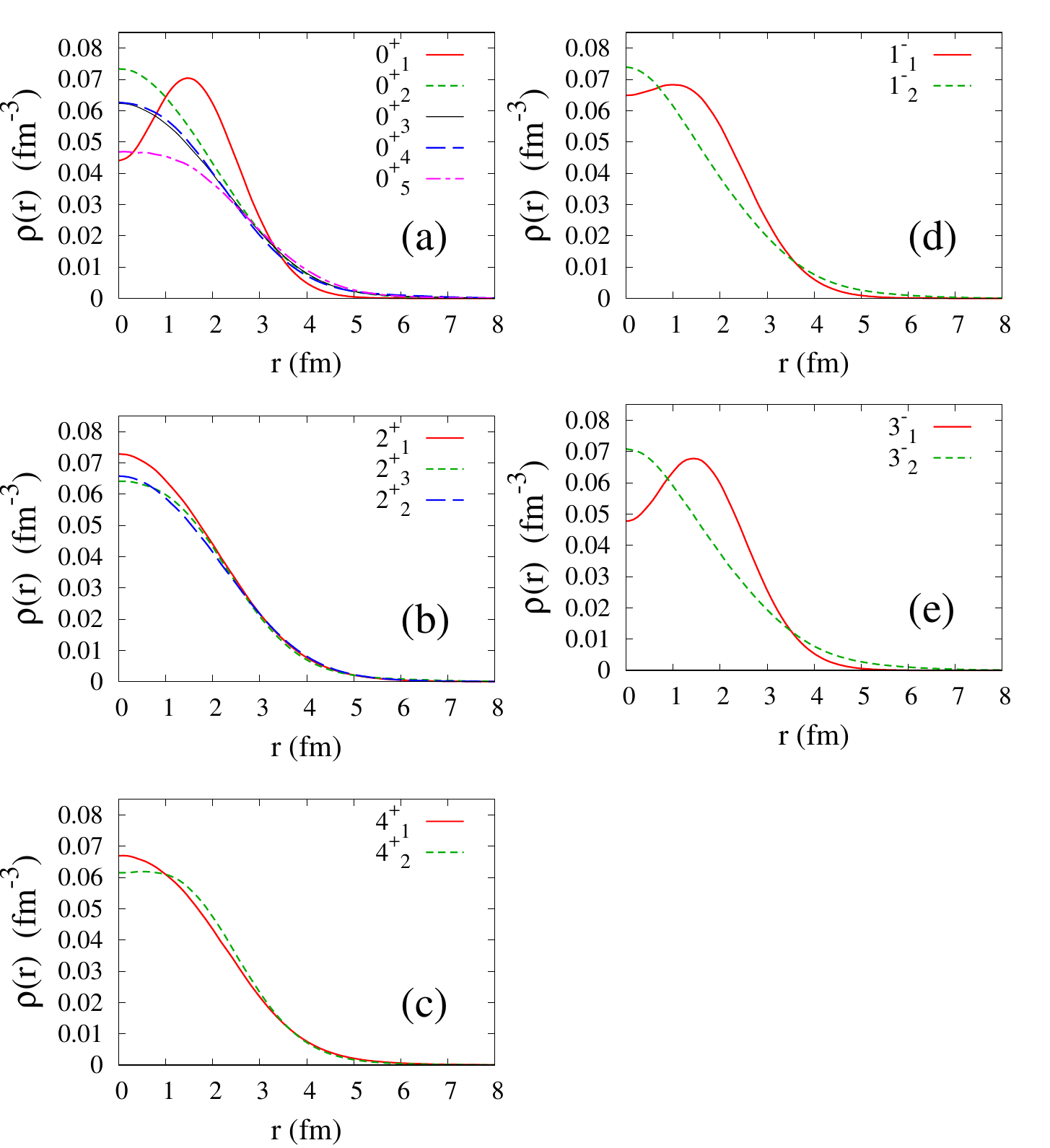}
\end{center}
  \caption{Proton density $\rho_p(r)=\rho(r)/2$
of the $0^+_{1,2,3,4,5}$, $1^-_{1,2}$, $2^+_{1,2,3}$, $3^-_{1,2}$, and
$4^+_{1,2}$ states of $^{16}$O calculated with the VAP+GCM.
  \label{fig:density}}
\end{figure}

\begin{table}[ht]
\caption{Excitation energies $E_x$ (MeV) and rms matter radii $R$ (fm) of $^{16}$O
calculated with the VAP+GCM. The experimental values of the excitation energies from
Ref.~\cite{Tilley:1993zz} are also shown.
The experimental data $R=2.55$~fm of the rms radius of the ground state is
deduced from the experimental charge radius measured by the electron scattering \cite{Angeli2013}.
We assign the
fourth $0^+$, third $0^+$, third $2^+$, and the second $2^+$ states obtained by the VAP+GCM
to the experimental
$0^+_{3}$,  $0^+_{4}$, $2^+_2$, and $2^+_3$ states, which we label as
$0^+_{3,\textrm{IV}}$,  $0^+_{4,\textrm{III}}$, $2^+_{2,\textrm{III}}$, and $2^+_{3,\textrm{II}}$, respectively.
 \label{tab:radii}
}
\begin{center}
\begin{tabular}{lrrrrrrrrccccc}
\hline
\hline
    & {exp}       & \multicolumn{2}{c}{VAP+GCM} \\
    & $E_x$ (MeV) &     $E_x$ (MeV) & $R$ (fm) \\
$0^+_1$ & 0 & 0.0   & 2.73  \\
$0^+_2$ & 6.0494  & 9.7   & 3.29  \\
$0^+_{3,\textrm{IV}}$ & 12.049  & 15.3  & 3.53  \\
$0^+_{4,\textrm{III}}$  & 13.6  & 13.6  & 3.64  \\
$0^+_5$ & 14.01 & 18.3  & 3.53  \\
$2^+_1$ & 6.917 & 10.8  & 3.27  \\
$2^+_{2,\textrm{III}}$  & 9.846 & 14.7  & 3.37  \\
$2^+_{3,\textrm{II}}$ & 11.52 & 14.0  & 3.51  \\
$2^+_4$ & 13.04 & 16.1  & 3.55  \\
$2^+_5$ & 14.926  & 16.9  & 3.26  \\
$2^+_6$ & 15.26 & 18.9  & 3.56  \\
$2^+_7$ & 16.44 & 20.3  & 3.86  \\
$2^+_8$ & 16.93 & 20.6  & 3.38  \\
$4^+_1$ & 10.356  & 13.7  & 3.34  \\
$4^+_2$ & 11.097  & 14.5  & 3.18  \\
$4^+_3$ & 13.869  & 16.6  & 3.63  \\
$1^-_1$ & 7.1169  & 9.4   & 2.87  \\
$1^-_2$ & 9.585   & 12.1  & 3.58  \\
$3^-_1$ & 6.1299  & 7.6   & 2.78  \\
$3^-_2$ & 11.600    & 13.4  & 3.65  \\
\hline
\end{tabular}
\end{center}
\end{table}

The excitation energies and radii of $^{16}$O from the VAP+GCM calculation and the experimental data
are listed in Table \ref{tab:radii}.
We assign the
fourth $0^+$($0^+_\textrm{IV})$, the third $0^+$($0^+_\textrm{III})$, the third $2^+$($2^+_\textrm{III})$, and the second $2^+$($2^+_\textrm{II})$ states in the theoretical spectra
to the experimental levels of the
$0^+_{3}$(12.05 MeV),  $0^+_{4}$(13.6 MeV), $2^+_2$(9.85 MeV), and $2^+_3$(11.52 MeV)
states, respectively,
because the  VAP+GCM calculation gives  incorrect ordering of the $K^\pi=0^+_3$ and $0^+_4$ bands;
for the detailed discussion, see Ref.~\cite{Kanada-En'yo:2013dma}.
The nuclear sizes of the $3^-_1$ and $1^-_1$ states are comparable
to the ground state and relatively smaller than those of other excited states because
the $3^-_1$ is the $4\alpha$-tetrahedral  state in the ground band and
the $1^-_1$ is the vibration mode on the  tetrahedral ground band.
The $4^+_2$ state is also regarded as the $4\alpha$-tetrahedral band but its size is slightly larger than those of the other two
in the ground band because of
the mixing with the $4^+_1$ state in the $^{12}$C+$\alpha$ cluster band.
Other states are developed cluster states and
have relatively larger radii than those of the ground-band states.
The density distribution of the $0^+_{1,2,3,4,5}$, $1^-_{1,2}$, $2^+_{1,2,3}$, $3^-_{1,2}$, and
$4^+_{1,2}$ states obtained by the VAP+GCM is shown in Fig.~\ref{fig:density}.
The $0^+_{2,3,4,5}$, $1^-_{2}$, $2^+_{1,2,3}$, $3^-_{2}$, and $4^+_1$ states
tend to have the slightly enhanced surface density in the range of $r=4$--5~fm
because of the developed cluster structures.

\subsection{Transition strengths, transition densities, and charge form factors of $^{16}$O}

\begin{table}[ht]
\caption{
The transition strengths $B(E\lambda)$ of $^{16}$O
calculated with the VAP+GCM and the experimental data from Refs.~\cite{Tilley:1993zz,Buti:1986zz}.
For the IS dipole transition strengths
of the $1^-\to 0^+$ transitions, the values of $B(\textrm{IS}1)/4$ are shown.
The scaling factor $f_\textrm{tr}=\sqrt{B_\textrm{exp}(E\lambda)/B_\textrm{cal}(E\lambda)}$
determined by the ratio of the experimental value $B_\textrm{exp}(E\lambda)$
to the calculated value $B_\textrm{cal}(E\lambda)$ for each transition is also shown.
For the transitions with no experimental data of $B(E\lambda)$,
$f_\textrm{tr}=1$ is used.
The units are $e^2$fm$^{2\lambda}$ for $B(E\lambda)$  ($\lambda\ne 0$),
$e^2$fm$^{4}$ for $B(E0)$, and fm$^{6}$ for $B(\textrm{IS}1)$.
$^a$The calculated $B(E2:2^+_2\to 0^+_2)$ is too small, and therefore, is not adjusted to the experimental value
but $f_\textrm{tr}=1$ is used.
$^b$For the scaling factor of the $1^-_1\to 0^+_1$ transition, we use $f_\textrm{tr}=1$ in the default CC calculation, and also use
the modified value $f_\textrm{tr}=1.3$, which are phenomenologically adjusted so as to reproduce the charge form factors and $\alpha$ scattering
cross sections.
 \label{tab:BEl}
}
\begin{center}
\begin{tabular}{cccccccccc}
\hline
& \multicolumn{2}{l}{exp} &\multicolumn{2}{l}{VAP+GCM}    \\
    &     $B(E\lambda)$ \cite{Tilley:1993zz} &  $(e,e')$  \cite{Buti:1986zz} &    $B(E\lambda)$ & $f_\textrm{tr}$ \\
$E2:2^+_1\to 0^+_1$ & 7.42  $(  0.24  )$  & 7.79  & 3.05  & 1.56  \\
$E2:2^+_1\to 0^+_2$ & 65  $(  7   )$  &   & 140   & 0.68  \\
$E2:2^+_2\to 0^+_1$ & 0.07  $(  0.01  )$  & 0.05  & 0.29  & 0.51  \\
$E2:2^+_2\to 0^+_2$ & 2.87  $(  0.72  )$  &   & 0.02  & 1$^a$ \\
$E2:2^+_3\to 0^+_1$ & 3.59  $(  1.20  )$  & 3.40  & 2.39  & 1.23  \\
$E2:2^+_3\to 0^+_2$ & 7.42  $(  1.20  )$  &   & 43.7  & 0.41  \\
$E2:4^+_1\to 2^+_1$ & 156   $(  14  )$  &   &   & 1 \\
$E2:4^+_2\to 2^+_1$ & 2.39  $(  0.72  )$  &   &   & 1 \\
$E2:1^-_1\to 3^-_1$ & 50  $(  12  )$  &   & 33.7  & 1.22  \\
$E2:1^-_2\to 3^-_1$ &         &   & 1.0   & 1 \\
$E0:0^+_2\to 0^+_1$ & 12.6        & 11.8  & 12.0  & 1.03  \\
$E0:0^+_3\to 0^+_1$ & 16.2        & 14.2  & 16.7  & 0.99  \\
$E0:0^+_4\to 0^+_1$ &         &   & 10.7  & 1\\
$E0:0^+_5\to 0^+_1$ & 10.9        &   & 9.0   & 1.10  \\
$E3:3^-_1\to 0^+_1$ & 205   $(  11  )$  &   & 207   & 0.99  \\
$E3:3^-_1\to 0^+_2$ &         &   &   & 1 \\
$E3:3^-_2\to 0^+_1$ &         &   &   & 1 \\
$E3:3^-_2\to 0^+_2$ &         &   &   & 1 \\
$E4:4^-_1\to 0^+_1$ & 378   $(  133   )$  & 420   & 345   & 1.10  \\
$E4:4^-_1\to 0^+_2$ &         &   &   & 1 \\
$E4:4^-_2\to 0^+_1$ &         & 372   & 71  & 2.30  \\
$E4:4^-_2\to 0^+_2$ &         &   &   & 1 \\
\\
IS1:$1^-_1\to  0^+_1$ &         &   & 13.8  & 1(1.3)$^b$  \\
IS1:$1^-_1\to  0^+_2$ &         &   & 74  & 1 \\
IS1:$1^-_2\to  0^+_1$ &         &   & 0.2   & 1 \\
IS1:$1^-_2\to  0^+_2$ &         &   & 745   &   1 \\
\hline
\end{tabular}
\end{center}
\end{table}

\begin{figure}[!h]
\begin{center}
\includegraphics[width=7cm]{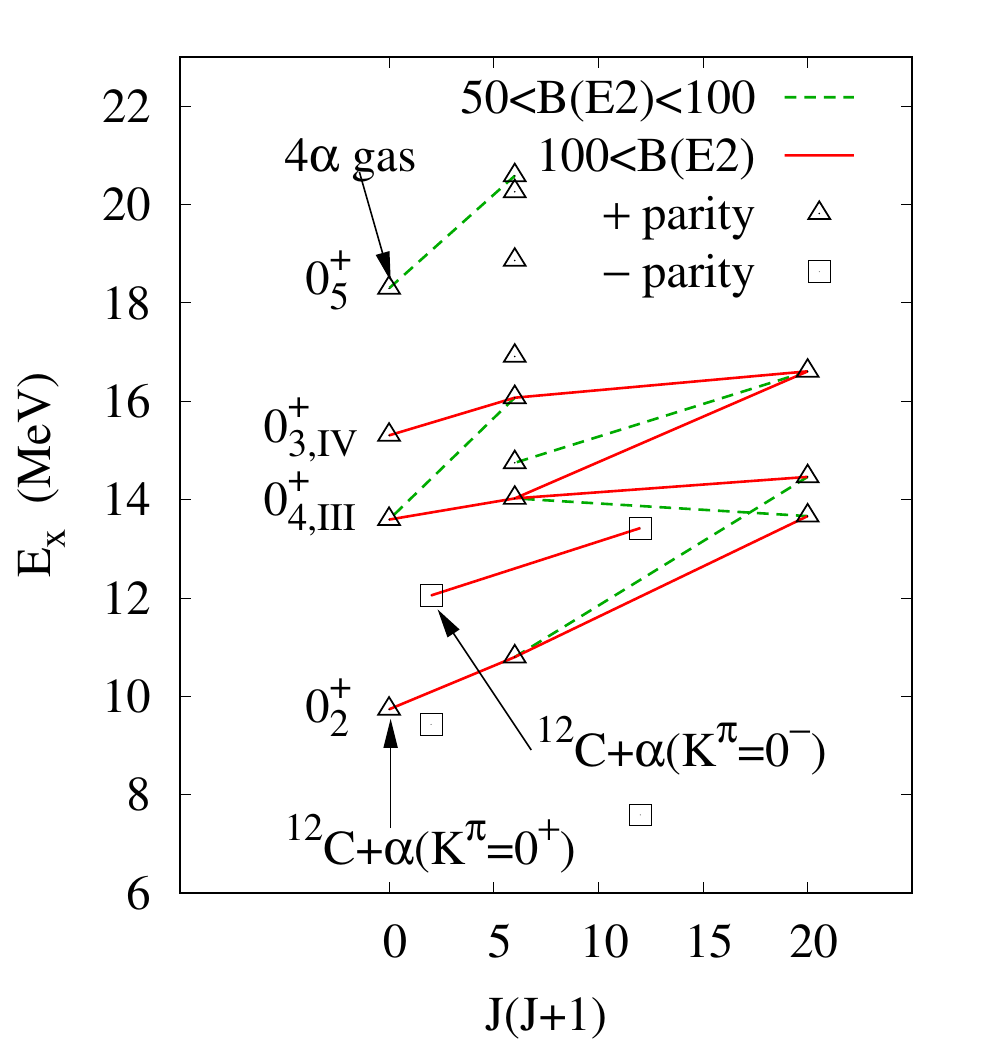}
\end{center}
  \caption{The energy levels adopted in the CC calculation and the $E2$ transitions.
The positive-parity states  ($0^+$, $2^+$, and $4^+$) are shown by open triangles  and the
negative-parity states  ($1^-$ and $3^-$) are shown by open squares.
The energy levels are connected by (green) dashed and (red) solid lines
in case of $50< B(E2)< 100$ $e^2$fm$^4$  and $B(E2)>100$ $e^2$fm$^4$, respectively.
  \label{fig:spe}}
\end{figure}

\begin{figure}[!h]
\begin{center}
\includegraphics[width=8.6cm]{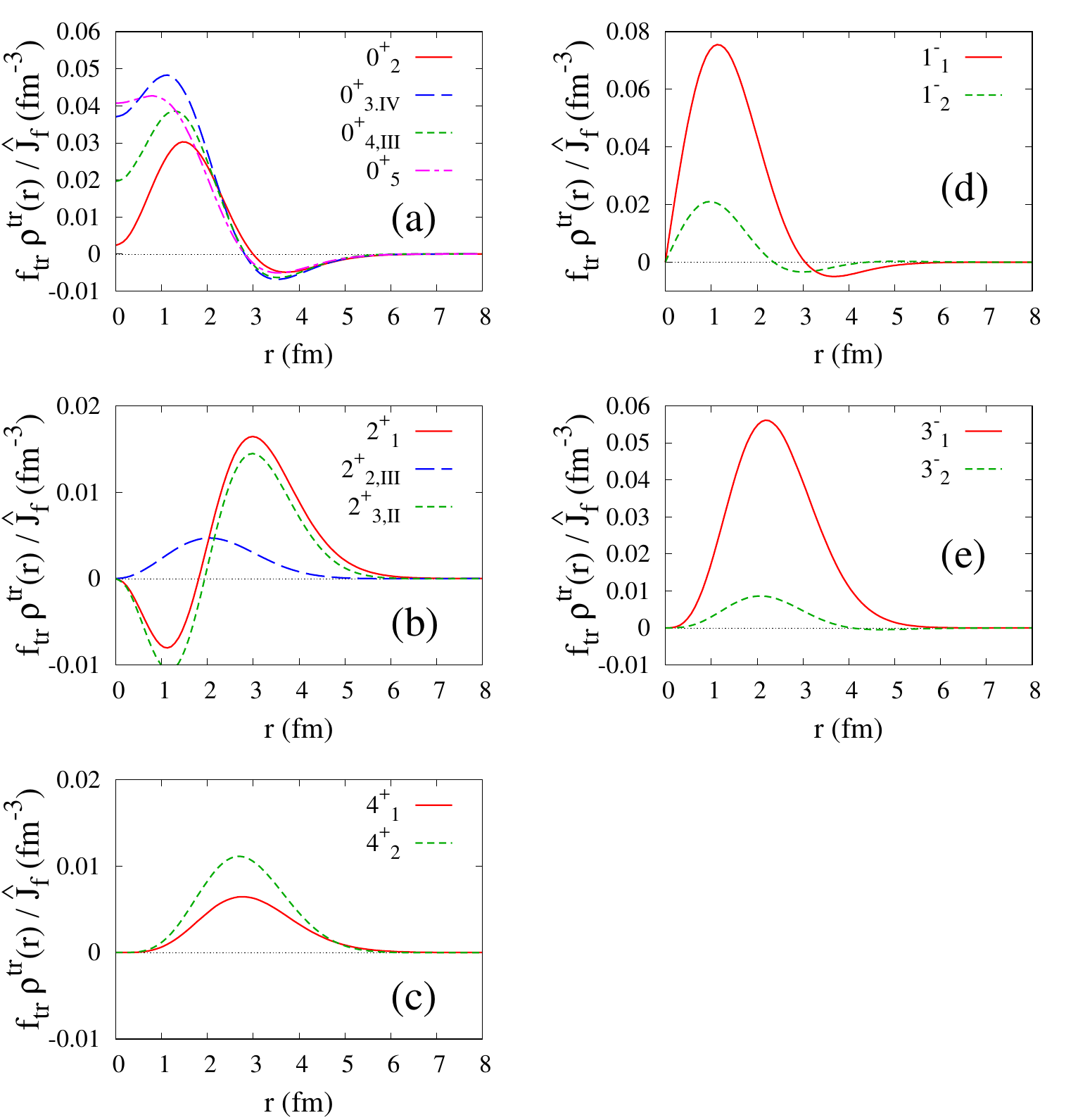}
\end{center}
  \caption{Transition density for the transitions from the ground state in
$^{16}$O calculated with the VAP+GCM. The scaled transition density $f_\textrm{tr} \rho^\textrm{(tr)}(r)$
divided by $\hat{J_f}\equiv=\sqrt{2 J_f+1}$
is shown. 
  \label{fig:transition}}
\end{figure}

\begin{figure*}[!h]
\begin{center}
\includegraphics[width=15cm]{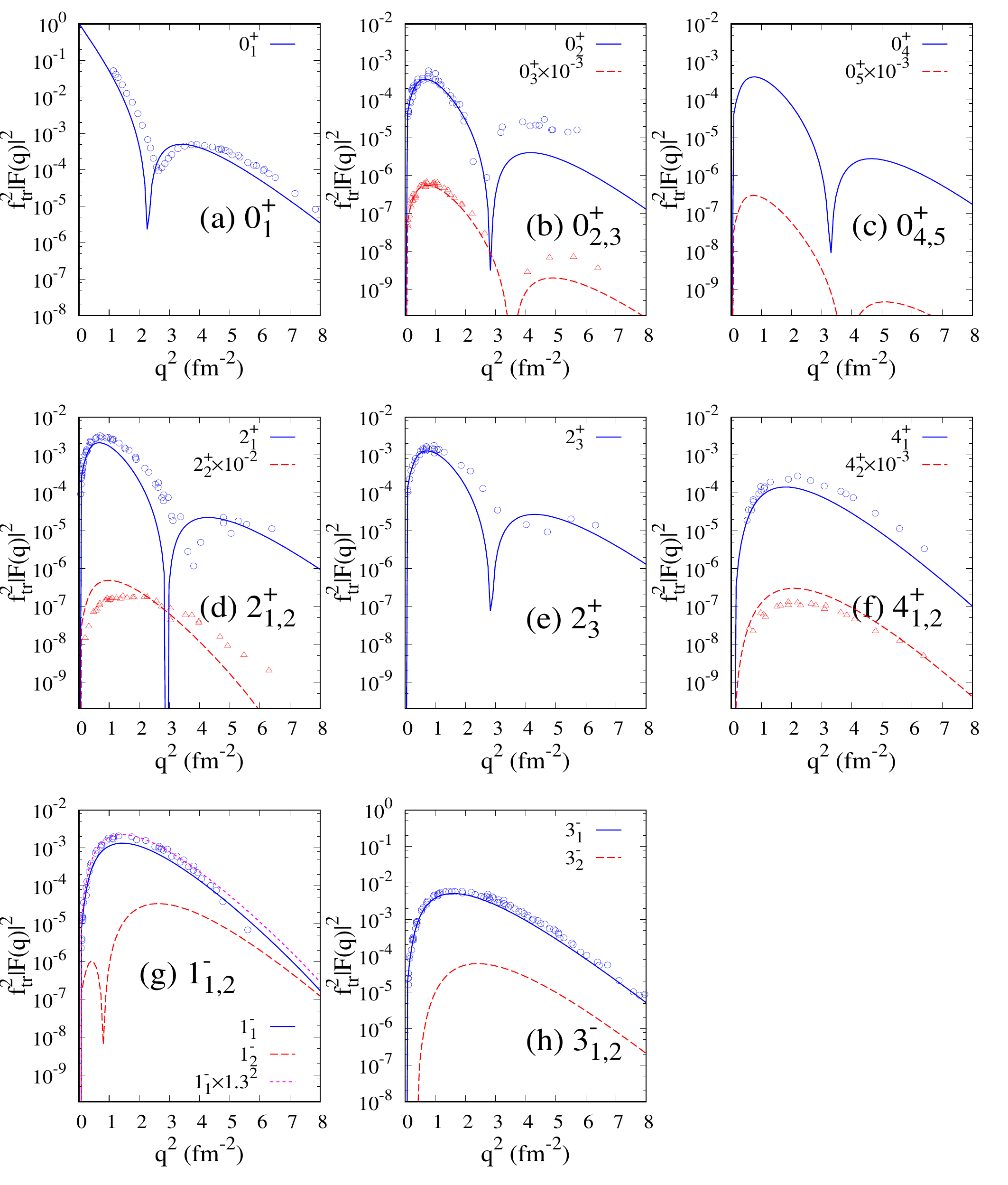}
\end{center}
  \caption{Squared charge form factors of $^{16}$O.
The theoretical values obtained with the VAP+GCM are scaled by $f^2_\textrm{tr}$ consistently with
the scaled transition density.
For the $1^-_1\to 0^+_1$ transition, the squared form factors scaled by
the modified value $f_\textrm{tr}^2=1.3^2$ of the scaling factor are also shown.
The experimental data are from Ref.~\cite{Buti:1986zz}.
  \label{fig:form}}
\end{figure*}

In Table \ref{tab:BEl},
the transition strengths, $B(E\lambda)$, of $^{16}$O
calculated with the VAP+GCM are shown
compared with the experimental data.
For the IS dipole transition strengths
of the $1^-\to 0^+$ transitions, the values of $B(\textrm{IS}1)/4$ are shown.
The energy levels and the major $E2$ transitions are shown in Fig.~\ref{fig:spe}. In the figure,
the energy levels are
connected by (green) dashed and (red) solid lines
for the case of remarkable $E2$ transitions of $50< B(E2)< 100$ $e^2$fm$^4$ and $B(E2)>100$ $e^2$fm$^4$,
respectively.
Rather strong $E2$ transitions are found for developed cluster states. Some of those remarkable
$\lambda=2$ transitions give significant CC effects to the $\alpha$ scattering cross sections as discussed later.
The $K^\pi=0^+_2$ band of the $^{12}$C+$\alpha$ cluster is composed of the
$0^+_2$, $2^+_1$, and $4^+_1$ states, and its parity-partner $K^\pi=0^-$ band is
constructed by the $1^-_2$ and $3^-_2$ states. The $4\alpha$-gas state is obtained as
the $0^+_5$ state.

To reduce ambiguity from the structure model calculation
in application of the theoretical transition density to the reaction calculation,
we scale the calculated result as $\rho^\textrm{(tr)}(r) \to f_\textrm{tr} \rho^\textrm{(tr)}(r)$
to fit the observed $B_\textrm{exp}(E\lambda)$. The scaling factor
$f_\textrm{tr}=\sqrt{B_\textrm{exp}(E\lambda)/B_\textrm{cal}(E\lambda)}$ introduced here is defined by
the square root of the ratio of the experimental  $B(E\lambda)$ value to the theoretical one.
The adopted value of $f_\textrm{tr}$ for each transition is shown in Table \ref{tab:BEl}.
For the transitions for which experimental data of $B(E\lambda)$ do not exist, we take $f_\textrm{tr}=1$ and
use the original transition density.
For the $1^-_1\to 0^+_1$ transition, $B(\textrm{IS1})$ is unknown, but the charge form factors are available from
the $(e,e')$ reaction data. For this transition, we use $f_\textrm{tr}=1$ in the default CC calculation, and also test
a modified value $f_\textrm{tr}=1.3$ of the $0^+_1\to 1^-_1$ transition density, which consistently reproduces the charge form factors and $\alpha$ scattering
cross sections.
Figure \ref{fig:transition} shows
the scaled transition density $f_\textrm{tr} \rho^\textrm{(tr)}(r)$ for transitions from the ground state.

We show in Fig.~\ref{fig:form}
the charge form factors of $^{16}$O calculated with the VAP+GCM  and
the experimental data measured by the electron scattering \cite{Buti:1986zz}.
The calculated form factors are scaled by multiplying $f^2_\textrm{tr}$ consistently with the scaled transition density.
The experimental data are
reasonably reproduced by the scaled form factors of the VAP+GCM.
In the $E2$ form factors of the $2^+$ states (Fig.~\ref{fig:form}(d)),
a clear difference can be seen in the $0^+_1\to 2^+_2$ from the other $E2$ transitions of $0^+_1\to 2^+_{1,3}$
because the corresponding transition density of $0^+_1\to 2^+_2$ shows
the different behavior that it is the compact spatial distribution with no nodal structure (see Fig.~\ref{fig:transition}(b)).
In the $E0$ form factors of $0^+$ states, the $0^+_1\to 0^+_2$ transition has
the dip at the smallest momentum transfer ($q^2$) corresponding to the broadest distribution
of the transition density.

\section{$\alpha$ scattering} \label{sec:scattering}
\subsection{Coupled-channel calculation}
Using the matter and transition densities calculated with the VAP+GCM
as the input from the structure calculation, we perform the CC
calculation of $^{16}$O($\alpha,\alpha')$ with the $g$-matrix folding model in the same way as in our previous work \cite{Kanada-En'yo:a12C}.
The $\alpha$-$^{16}$O CC potentials are constructed
by folding the Melbourne $g$-matrix $NN$ interaction \cite{Amos:2000} with the densities of $\alpha$ and
$^{16}$O in the approximation of an extended version of the nucleon-nucleus folding (NAF) model \cite{Egashira:2014zda}. For the $\alpha$ density, we adopt the one-range Gaussian distribution given in Ref.~\cite{Satchler:1979ni}.

In the default CC calculation of the cross sections of the
$0^+$, $1^-$, $2^+$, and $3^-$ states,
we adopt the $0^+_{1,2,3,4,5}$, $2^+_{1,2,3,4}$, $1^-_{1,2}$, and $3^-_{1,2}$ states with the
$\lambda \le 3 $ transitions in the target $^{16}$O nucleus. The scaled transition density
$f_\textrm{tr} \rho^\textrm{(tr)}(r)$ is used.
For the excitation energies of  $^{16}$O, we use the experimental values listed in Table~\ref{tab:radii}.
In the calculation of the $4^+$ cross sections, we adopt
$0^+_{1,2,3,4}$, $2^+_{1,2,3,4,5,6,7,8}$, and $4^+_{1,2,3}$ states with the $\lambda=0,2,4$ transitions.
For the cross sections to the $0^+_{5}$ state, we also perform the CC calculation using the
$0^+_{1,2,3,4,5}$ and $2^+_{1,2,3,4,5,6,7,8}$ states with the $\lambda=0,2$ transitions
to take into account the strong $E2$ transition between the $0^+_5$ and
$2^+_{8}$ states and compare the result with the default CC calculation.

\subsection{$\alpha$ scattering cross sections}

\begin{figure*}[!h]
\begin{center}
\includegraphics[width=17cm]{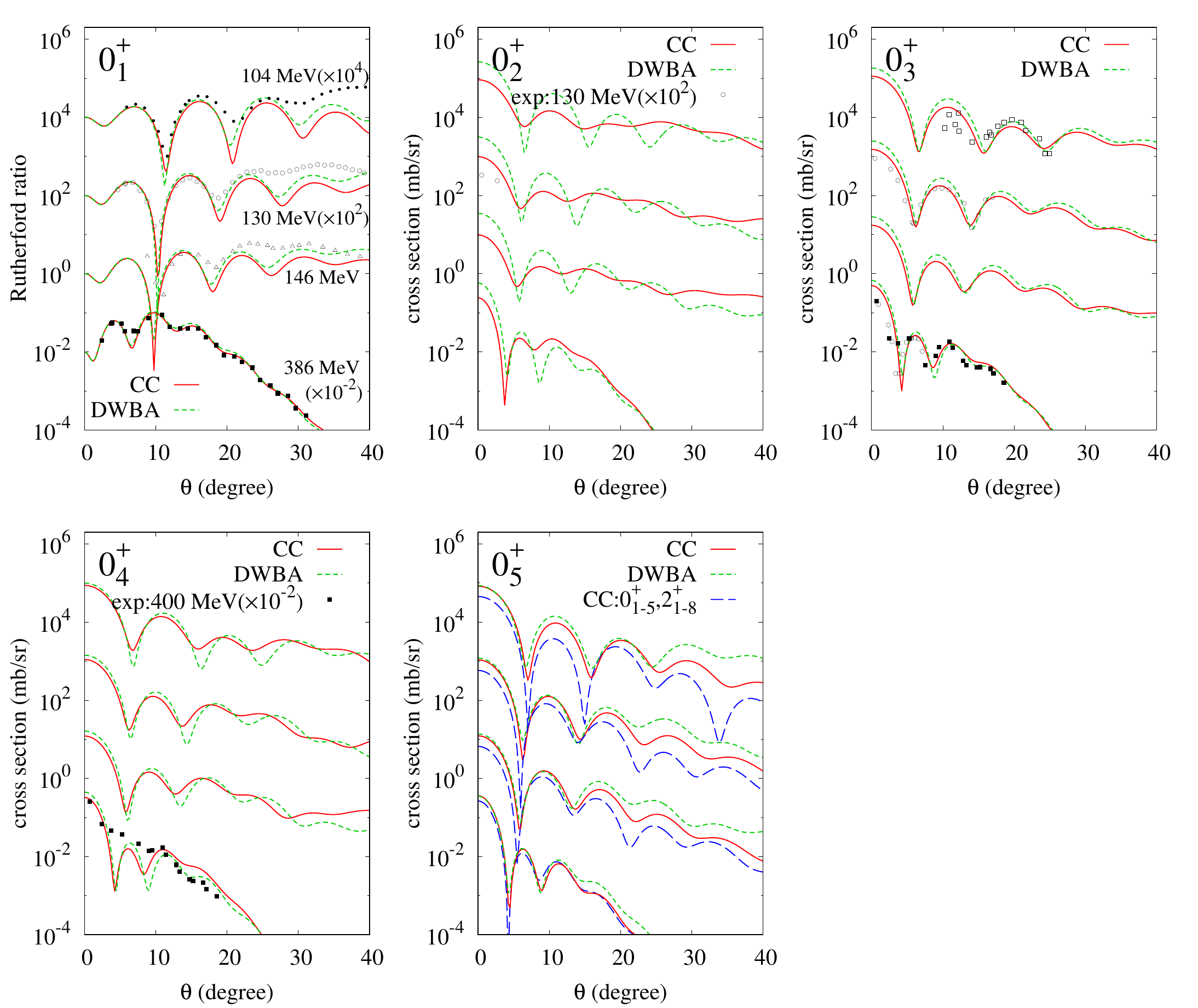}
\end{center}
  \caption{
$\alpha$ scattering cross sections
on $^{16}$O at $E_\alpha=104$ MeV ($\times 10^4$), 130 MeV  ($\times 10^2$), 146 MeV, and 386 MeV  ($\times 10^{-2}$).
The differential cross sections of the $0^+_{1,2,3,4,5}$
obtained by the CC and DWBA calculations are shown by (red) solid and (green) dashed lines, respectively.
For the $0^+_{5}$, the cross sections obtained by the CC calculation using the  $0^+_{1,2,3,4,5}$ and $2^+_{1,2,\ldots,8}$ states
are also shown by (blue) long-dashed lines.
The experimental data at $E_\alpha=104$ MeV \cite{Hauser:1969hkb}, 130 MeV \cite{Adachi:2018pql},  146 MeV \cite{Knopfle1975}, and 400 MeV \cite{Wakasa:2006nt} are shown by filled circles, open circles, open triangles, and filled squares, respectively.
For the $0^+_3$ cross sections, the data at $E_\alpha=104$ MeV from Ref.~\cite{Harakeh:1976rtd} and those
at $E_\alpha=386$ MeV from Ref.~\cite{Adachi:2018pql} are also shown by open squares and open circles, respectively.
\label{fig:cross-J0}}
\end{figure*}

\begin{figure*}[!h]
\begin{center}
\includegraphics[width=17cm]{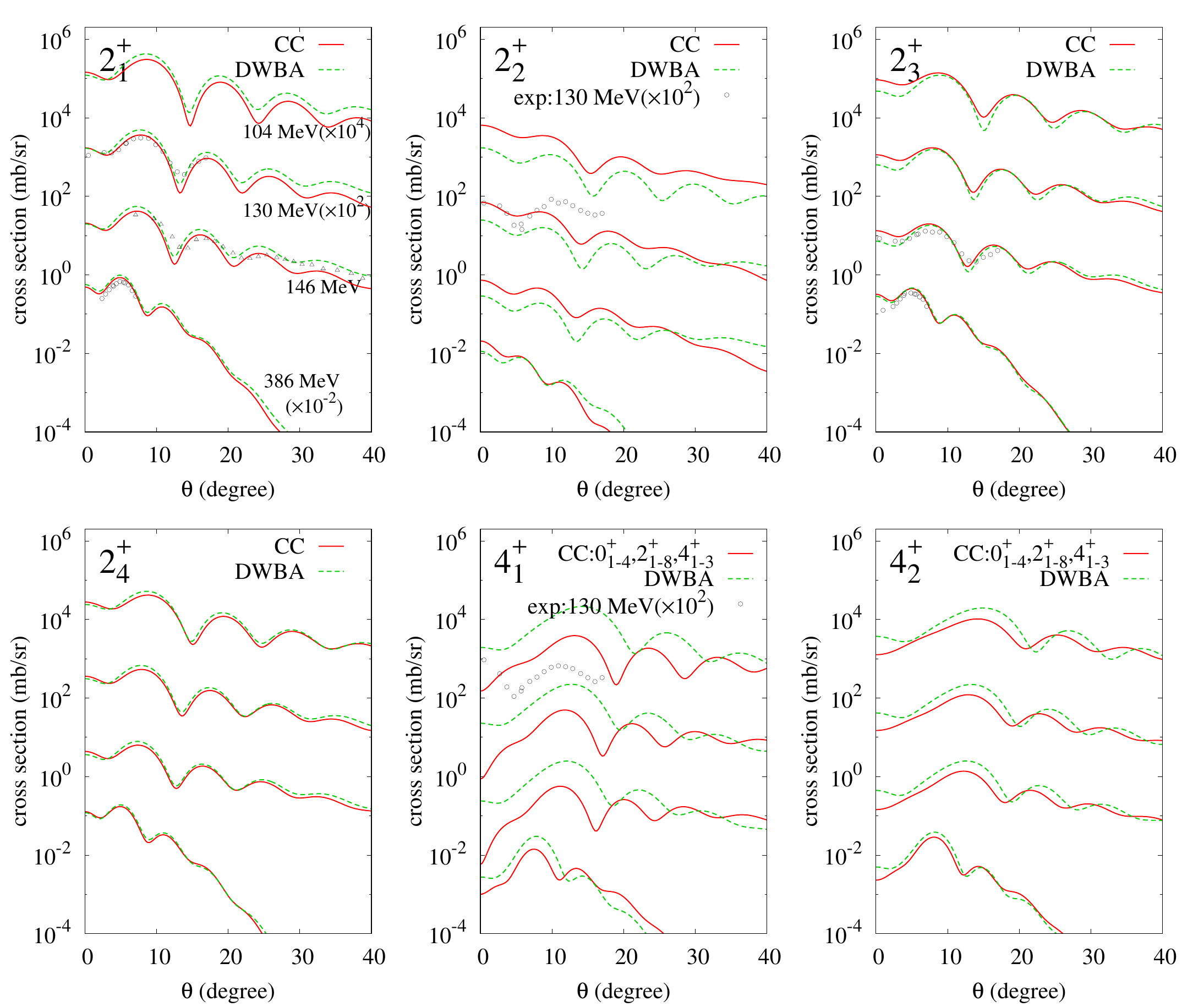}
\end{center}
  \caption{$\alpha$ scattering cross sections
on $^{16}$O at $E_\alpha=104$ MeV ($\times 10^4$), 130 MeV  ($\times 10^2$), 146 MeV, and 386 MeV  ($\times 10^{-2}$).
The differential cross sections of the $2^+_{1,2,3,4}$ and $4^+_{1,2}$
obtained by the CC and DWBA calculations are shown
by (red) solid and (green) dashed lines.
In the CC calculation for the $4^+_{1,2}$ cross sections,
the $0^+_{1,2,3,4}$, $2^+_{1,2,\ldots,8}$, and
$4^+_{1,2,3}$ states are used.
The experimental data at 130 MeV \cite{Adachi:2018pql},  146 MeV \cite{Knopfle1975}, and 386 MeV \cite{Adachi:2018pql} are shown
by open circles, open triangles, and open circles, respectively.
  \label{fig:cross-J2}}
\end{figure*}

\begin{figure*}[!h]
\begin{center}
\includegraphics[width=17cm]{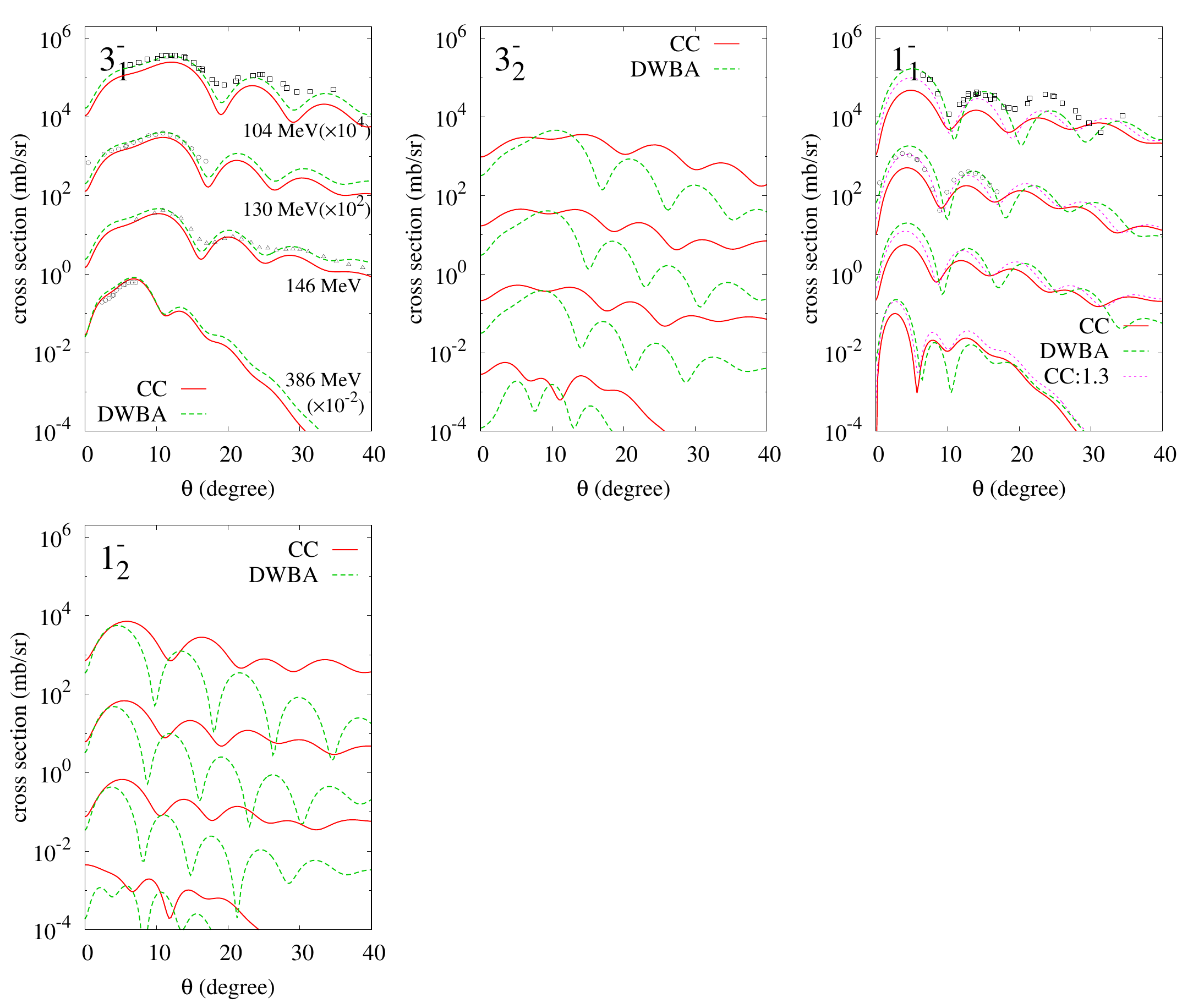}
\end{center}
  \caption{$\alpha$ scattering cross sections
on $^{16}$O at $E_\alpha=104$ MeV ($\times 10^4$), 130 MeV  ($\times 10^2$), 146 MeV, and 386 MeV  ($\times 10^{-2}$).
The differential cross sections of the $3^-_{1,2}$ and $1^-_{1,2}$
obtained by the default CC and DWBA calculations are shown by (red) solid and (green) dashed lines.
The experimental data at 104 MeV \cite{Harakeh:1976rtd}, 130 MeV \cite{Adachi:2018pql},  146 MeV \cite{Knopfle1975}, and 386 MeV \cite{Adachi:2018pql} are shown
by open squares, open circles, open triangles, and open circles, respectively. For the $1^-_1$ states, the result calculated
with the modified value
$f_\textrm{tr}=1.3$ of the scaling factor for the $0^+_1\to 1^-_1$ transition density are also shown by (magenta) dotted lines.
  \label{fig:cross-J3}}
\end{figure*}

The $\alpha$ scattering cross sections at the incident energies of $E_\alpha=104$~MeV, 130~MeV, 146~MeV, and 386~MeV
are shown in Figs.~\ref{fig:cross-J0}, \ref{fig:cross-J2},
and \ref{fig:cross-J3}, respectively.
The cross sections calculated by the DWBA are also shown for comparison.

The elastic cross sections  are well reproduced by the present
calculation except at large scattering angles for $E_\alpha$=104--146~MeV.
For the $\lambda=2$ and $\lambda=3$ transitions to the $2^+_1$, $2^+_3$, and $3^-_1$ states,
the calculated cross sections are in good agreement with the experimental data.
These states are strongly populated in the direct transitions, and the cross sections are dominantly described by the
DWBA calculation. For these states, the CC effect is minor, in particular, at $E_\alpha$=386~MeV,
but not negligible in the cross sections at the relatively low incident energies of  $E_\alpha$=104--146~MeV.
For the $2^+_2$, the calculation reproduces the absolute amplitude of the cross sections at forward angles but
does not satisfactorily describe the diffraction pattern of the observed data.
For the $4^+_1$ state, the present calculation predicts a very weak population and much underestimates
the experimental cross sections.

For the IS dipole excitation to the $1^-_1$ state, the experimental cross sections are somewhat underestimated
by the default CC calculation (solid lines of Fig.~\ref{fig:cross-J3}) with the original $1^-_1\to 0^+_1$
transition density, but successfully reproduced by
the calculation with the modified $0^+_1\to 1^-_1$ transition density scaled by the
factor of $f_\textrm{tr}=1.3$, which reproduces the charge form factors of this transition.
In comparison with the DWBA calculation,
one can see the significant CC effect in the $1^-_1$ cross sections.
Namely, the absolute amplitude of the cross sections is
drastically reduced and the dip positions are slightly shifted to forward angles.
This CC effect, which is mainly through the $3^-_1$ state, is essential to reproduce
the first dip position of the experimental cross sections at $E_\alpha=104$ MeV and 130 MeV.

For the monopole excitations, the calculated  $0^+_3$ and $0^+_4$ cross sections are in good agreement with the
observed data. The present CC calculation describes well not only
the diffraction pattern but also the absolute amplitude in the wide range of the incident energies of $E_\alpha=104$--386~MeV,
and there is no overshooting problem of the $0^+$ cross sections.

In comparison with the DWBA calculation, one can see how the CC effect contributes to the monopole transitions
in the $\alpha$ scattering. In the $0^+_3$ and $0^+_4$ cross sections, the CC effect is not so large but not negligible, in particular,
at the low incident energies, $E_\alpha=104$--146 MeV. By contrast,
the CC effect gives the drastic change of the $0^+_2$ cross sections mainly because of the strong
in-band $E2$ transition with the $2^-_1$ state in the $^{12}$C+$\alpha$-cluster band built on the
$0^+_2$ state. At $E_\alpha=104$--146~MeV,
the peak amplitude of the
$0^+_2$ cross sections is largely reduced by about a factor of three from the result of the DWBA calculation.
Even at $E_\alpha=386$ MeV, the peak amplitude of the CC result is smaller by a factor of two than of DWBA.
Also for the $0^+_5$ cross sections, the CC effect is found to be of importance, because of the
strong $E2$ transitions between the $0^+_5$ and $2^+_8$ states with the developed cluster structure.
Although the CC effect seems to be not so strong in the default CC calculation without the coupling with the higher $2^+$ states
(the (red) solid lines in Fig.~\ref{fig:cross-J0}),
the CC calculation with the $0^+_{1,2,3,4,5}$ and $2^+_{1,2,\ldots,8}$ states shows the drastic CC effect of the
significant reduction of the  $0^+_5$ cross sections at the low
incident energies of $E_\alpha=104$--146 MeV (the (blue) long-dashed lines in Fig.~\ref{fig:cross-J0}).
The CC effect in the  $0^+_5$  cross sections becomes weak at the relatively high incident energy of $E_\alpha=386$ MeV.

In the experimental studies of the monopole transitions,
the $\alpha$ scattering cross sections have been used to deduce the monopole strengths based on the
reaction model analysis mainly with the DWBA calculation by
naively expecting the scaling law of the $\alpha$-scattering cross sections and the electric monopole transition strength, $B(E0)$.
However, the present analysis of the $\alpha$ scattering indicates that
the scaling law
is not necessarily valid for the cluster states. Firstly, the amplitude of the $0^+$ cross sections can be
significantly affected by the CC effect mainly through the strong $\lambda=2$ transitions between the developed
cluster states. Secondly, the scaling law is not satisfied even in the one-step process of the DWBA cross sections
because of the difference in the matter and transition densities between excited $0^+$ states. 
These results indicate that, for study of the monopole transitions by means of the $(\alpha,\alpha')$ reaction,
it is necessary to analyze the $\alpha$ scattering cross sections with
microscopic reaction models considering such the CC effect and density profiles.
Nevertheless, we should remark that $0^+$ cluster states with significant monopole strengths are
strongly populated by the $(\alpha,\alpha')$ reaction, meaning that it is still a good probe for the cluster states and
can be useful for qualitative discussion even though the scaling law is not quantitatively valid.

\section{Summary}\label{sec:summary}

The $\alpha$ inelastic scattering cross sections on $^{16}$O was investigated by
the folding model with the Melbourne $g$-matrix $NN$ interaction.
This is the first microscopic CC calculation of the
$^{16}$O($\alpha,\alpha')$ reaction that is based on the
$\alpha$-nucleus CC potentials microscopically derived with
the $g$-matrix $NN$ interactions and the matter and transition densities
of the target $^{16}$O nucleus calculated with the microscopic structure model.
As for the structure model,
we employed the VAP+GCM in the framework of the AMD, which is the microscopic approach beyond the
cluster models.
In the application to the reaction calculation,
the calculated transition density is scaled to fit the experimental transition strengths
to reduce the ambiguity of the structure model.

The calculation reproduces well the observed
cross sections of the $0^+_{2,3,4}$, $2^+_1$, $1^-_1$, and $3^-_1$ states as well as the elastic cross sections
at incident energies of
$E_\alpha=104$ MeV, 130 MeV, 146 MeV, and 386 MeV.
In the $0^+$ cross sections, there is no overshooting problem.
In comparison with the DWBA calculation,
the significant CC effect was found in the  $0^+_2$, $0^+_5$, and $1^-_1$ cross sections
because of the strong $\lambda=2$ coupling between excited states that have developed cluster structures.
We clarified that the scaling law
of the $\alpha$-scattering cross sections and $B(E0)$
is not necessarily satisfied for the cluster states because of the significant CC effect
through the strong $\lambda=2$ transitions between the developed
cluster states.
Nevertheless, it should be remarked that
the $(\alpha,\alpha')$ reaction can be used
for qualitative discussion on the cluster states because
$0^+$ cluster states with significant monopole strengths are
strongly populated by the $(\alpha,\alpha')$ reaction.

It is suggested that
the microscopic reaction calculation is needed
in the quantitative analysis of the $\alpha$ scattering cross sections.
The present $g$-matrix folding model was proved to be applicable to describe
the $\alpha$ scattering cross sections. This approach is a promising tool to extract
information on cluster structures of excited states in other nuclei by the $(\alpha,\alpha')$ reaction.

\begin{acknowledgments}
The computational calculations of this work were performed by using the
supercomputer in the Yukawa Institute for theoretical physics, Kyoto University. This work was supported
in part
by Grants-in-Aid of the Japan Society for the Promotion of Science (Grant Nos. JP18K03617, and JP16K05352) and by the grant for the RCNP joint research project.
\end{acknowledgments}

\end{document}